# Fano plasmonics goes nonlinear

Maxim Sukharev[1,2], Elena Drobnyh[2], Ruth Pachter[3]

[1]College of Integrative Sciences and Arts, Arizona State University, Mesa, Arizona 85212, USA
[2]Department of Physics, Arizona State University, Tempe, Arizona 85287, USA
[3]Air Force Research Laboratory, Materials and Manufacturing Directorate, Wright-Patterson Air Force Base, Dayton, Ohio 45433, USA

**Abstract**: We investigate the process of the second harmonic generation by plasmonic nano-antennas that exhibit Fano-like resonances. A rigorous fully vectorial Maxwell-hydrodynamics approach is employed to directly calculate the second order susceptibilities as function of the pump frequency, considering a periodic array of nanodolmens comprised of three Au nanorods. The results of the numerical simulations demonstrate noticeable enhancement of the second harmonic efficiency by the antisymmetric mode. Additionally, a simple analytical model based on two coupled nonlinear oscillators is proposed. It is shown that the second order optical response can be significantly enhanced at the frequency of the antisymmetric normal mode thus supporting our numerical results.

**Introduction**

The research in nanoplasmonics has attracted appreciable interest from various parts of the scientific community including physics,[1] materials science,[2] chemistry,[3] and biology.[4] At the heart of the primary interest in plasmonics is the strong electromagnetic field localization at resonant frequencies corresponding to surface plasmon-polariton modes.[5] Due to remarkable progress in nanomanufacturing, experimental capabilities have achieved tremendous spatial precision when building plasmonic systems ranging from nanoparticles of various shapes to metasurfaces comprised of periodic arrays of nanoparticles and/or nanoholes of different geometry.[6] Even though characteristic quality factors of plasmon modes are relatively low, large local field enhancements make such systems attractive for numerous applications in chemistry[7] and biology.[8] Furthermore, the strong field localization at resonant frequencies results in small mode volumes, which in turn opened up a new research direction now commonly referred to as polaritonic chemistry.[9] Here by using plasmonic systems as resonant cavities one can investigate how optics of quantum emitters (such as molecular aggregates, quantum dots, transition metal dichalcogenide monolayers, etc.) changes.[10] When the coupling strength between ensembles of quantum emitters and a local electromagnetic field surpasses all the damping rates, the system enters the strong coupling regime forming polaritonic states, which have properties of both light and matter.[11, 12] In addition to various fundamental questions arising from modeling, it has been shown that such materials can lead to modified properties.[12]

In recent years the field of nanoplasmonics has also entered the nonlinear regime.[13] A handful of investigations have been performed to quantify various nonlinear phenomena, ranging from second/third harmonic generation[14] through four-wave mixing[15] to optical bistability.[16] Moreover, coherent control has been applied to examine second harmonic generation by gold nanoparticles.[17] Such control combines strong spatial localization of electromagnetic fields with laser pulse shaping and ultrafast optical detection techniques that are now offering riveting time resolution as short as 2 fs.[18] It has also been shown first theoretically[19, 20] and then confirmed experimentally[21] that strongly coupled systems exhibit unique second harmonic generation signals demonstrating that polaritonic states directly participate in nonlinear dynamics.

Designing plasmonic materials for various applications in both linear and nonlinear regimes has been a subject of intense research. In particular, plasmonic systems that exhibit Fano-like resonances were proposed for induced transparency[22] and further local field enhancement.[23, 24] In this work we apply a numerical methodology based on the Maxwell-hydrodynamics equations to quantify the second harmonic generation for a Fano type plasmonic system, namely an array of nanodolmens.[25] Using a simple analytical model, we explain that such systems can lead to noticeable second harmonic enhancement when driven at a frequency of an antisymmetric normal mode.

**Second harmonic generation by gold nanodolmens**

It has been demonstrated experimentally[26] that Fano-type arrays comprised of gold nanopillar dimers can enhance second harmonic efficiency. The primary source of the enhancement was due to high local fields in the gap of each dimer. Recently other comprehensive computational studies[27] confirmed the second harmonic enhancement in asymmetric homodimers. In this work we concentrate on the physics of a Fano-type response and examine the simplest case of the bright-dark mode coupling, scrutinizing the second nonlinear response and its possible enhancement mechanism beyond the local field enhancement.

The system considered corresponds to a nanodolmen comprised of three Au nanorods arranged as schematically depicted in Fig. 1a. When irradiated by an incident field (linearly polarized in $x$-direction and propagating into the screen of the figure) the longitudinal mode of the horizontal nanorod is excited as illustrated in Fig. 1a. This is a bright mode that is directly excited by the incident field. Two parallel nanorods aligned in the $y$-direction support a dark mode corresponding to out-of-phase oscillations of the conduction electrons. When placed in close proximity of the horizontal rod such a mode can be excited by the bright mode due to the near field coupling. The coupling strength between the bright and dark modes and the frequency of the dark mode is controlled by geometrical parameters $G_1$ and $G_2$ (Fig. 1a).[25] In this manuscript we consider an experimentally realizable setup[28] arranging nanodolmens in a square lattice placed on a semi-infinite glass substrate. From experimental point of view, optical detection of the second harmonic signal from periodic systems is somewhat an easier task compared to dark field microscopy, for instance. Thus, the main effect considered below would be possible to observe in periodic systems. Our primary goal is to investigate the influence of Fano-type modes on the second harmonic generation. In principle, periodic plasmonic systems also support lattice modes[29] that may exhibit Fano lineshapes. However, in this paper we are interested in local Fano-like resonances supported by a single nanodolmen and considerations of how surface plasmon lattice modes may alter the second order response are beyond the scope of this work.

To model the optical response of the array we numerically solve Maxwell's equations

$$\dot{\mathbf{B}} = -\nabla \times \mathbf{E},$$
$$\varepsilon_0 \dot{\mathbf{E}} = \frac{1}{\mu_0} \nabla \times \mathbf{B} - \dot{\mathbf{P}}, \qquad (1)$$

where the macroscopic polarization, $\mathbf{P}$, satisfies the equation of motion following from the semi-classical hydrodynamics model[30]

$$\ddot{\mathbf{P}} + \gamma_e \dot{\mathbf{P}} = \frac{n_0 e^2}{m_e^*} \mathbf{E} + \frac{e}{m_e^*}\left(\dot{\mathbf{P}} \times \mathbf{B} - \mathbf{E}(\nabla \cdot \mathbf{P}) - \nabla p\right) - \frac{1}{n_0 e}\left((\nabla \cdot \dot{\mathbf{P}})\dot{\mathbf{P}} + (\dot{\mathbf{P}} \cdot \nabla)\dot{\mathbf{P}}\right),$$
$$\nabla p = \frac{5}{3}\frac{E_F}{e}\nabla(\nabla \cdot \mathbf{P}) - \frac{10}{9}\frac{E_F}{e^2 n_0}(\nabla \cdot \mathbf{P})\nabla(\nabla \cdot \mathbf{P}) - \frac{5}{27}\frac{E_F}{e^3 n_0^2}(\nabla \cdot \mathbf{P})^2 \nabla(\nabla \cdot \mathbf{P}). \qquad (2)$$

Here we also include the gradient of the electron pressure term, $\nabla p$, retaining all the terms up to the third order.[31] Equation (2) can also be expressed in nondimensional form similar to the Navier-Stokes equation. We introduce scaling parameters pertaining to characteristic length, $\lambda_0$, frequency, $\omega_0$, and peak amplitude of the electric field, $E_0$. It should be noted that the scaling of the macroscopic polarization cannot be written in a general form since $\mathbf{P}$ is a complex functional of the electric field, $\mathbf{E}$. Its nondimensional form, $\mathbf{P}'$, nevertheless can be expressed using characteristic length as follows $\mathbf{P} = n_0 e \lambda_0 \mathbf{P}'$. Equation (2) takes the form

$$\ddot{\mathbf{P}}' + A \cdot \dot{\mathbf{P}}' = B \cdot \left(\mathbf{E}' - \mathbf{E}'(\nabla \cdot \mathbf{P}')\right) + C \cdot \dot{\mathbf{P}}' \times \mathbf{B}' - \left((\nabla \cdot \dot{\mathbf{P}}')\dot{\mathbf{P}}' + (\dot{\mathbf{P}}' \cdot \nabla)\dot{\mathbf{P}}'\right) -$$
$$-\frac{5}{3}D \cdot \left(\nabla - \frac{2}{3}(\nabla \cdot \mathbf{P}')\nabla - \frac{1}{9}(\nabla \cdot \mathbf{P}')^2 \nabla\right)(\nabla \cdot \mathbf{P}'), \qquad (3)$$

where nondimensional coefficients read

$$A = \frac{\gamma}{\omega_0}, \quad B = \frac{eE_0}{m_e^* \lambda_0 \omega_0^2}, \quad C = \frac{eE_0}{m_e^* c \omega_0}, \quad D = \frac{E_F}{m_e^* \lambda_0^2 \omega_0^2}, \qquad (4)$$

and $\mathbf{E}' = \mathbf{E}/E_0$, $\mathbf{B}' = c\mathbf{B}/E_0$.

The following parameters are used to simulate gold: $n_0 = 5.90 \times 10^{28}$ m$^{-3}$, $m_e^* = 1.66 \times m_e$, $\gamma_e = 0.181$ eV, $E_F = 5.53$ eV. The numerical procedure employed is outlined in Supplementary Information.

The coupled equations (1) and (2) are discretized in space and time and numerically propagated using the finite-difference time-domain methodology.[32] The spatial resolution used is 1.5 nm and the time step is 2.5 attoseconds. The simulation domain spans 405 nm in the $x$ and $y$ directions, along which the periodic boundary conditions are applied, and 1.44 μm in the direction of the incident field propagation, resulting in total of $7 \times 10^7$ grid points. The absorbing boundary conditions (namely, convolutional perfectly matched layers[33]) are imposed in the longitudinal direction to absorb outgoing radiation. The home-built codes utilize the finite difference discretization of the Maxwell equations for efficient parallel simulations. The simulation domain is split into sub-domains with send/receive operations applied at all 6 faces of each sub-domain. Simulations are performed on a Cray XC40/50 ERDC DSRC cluster Onyx. Typical execution times vary between 20 minutes to obtain linear absorption spectra and 2 and a half hours to simulate nonlinear power spectra.

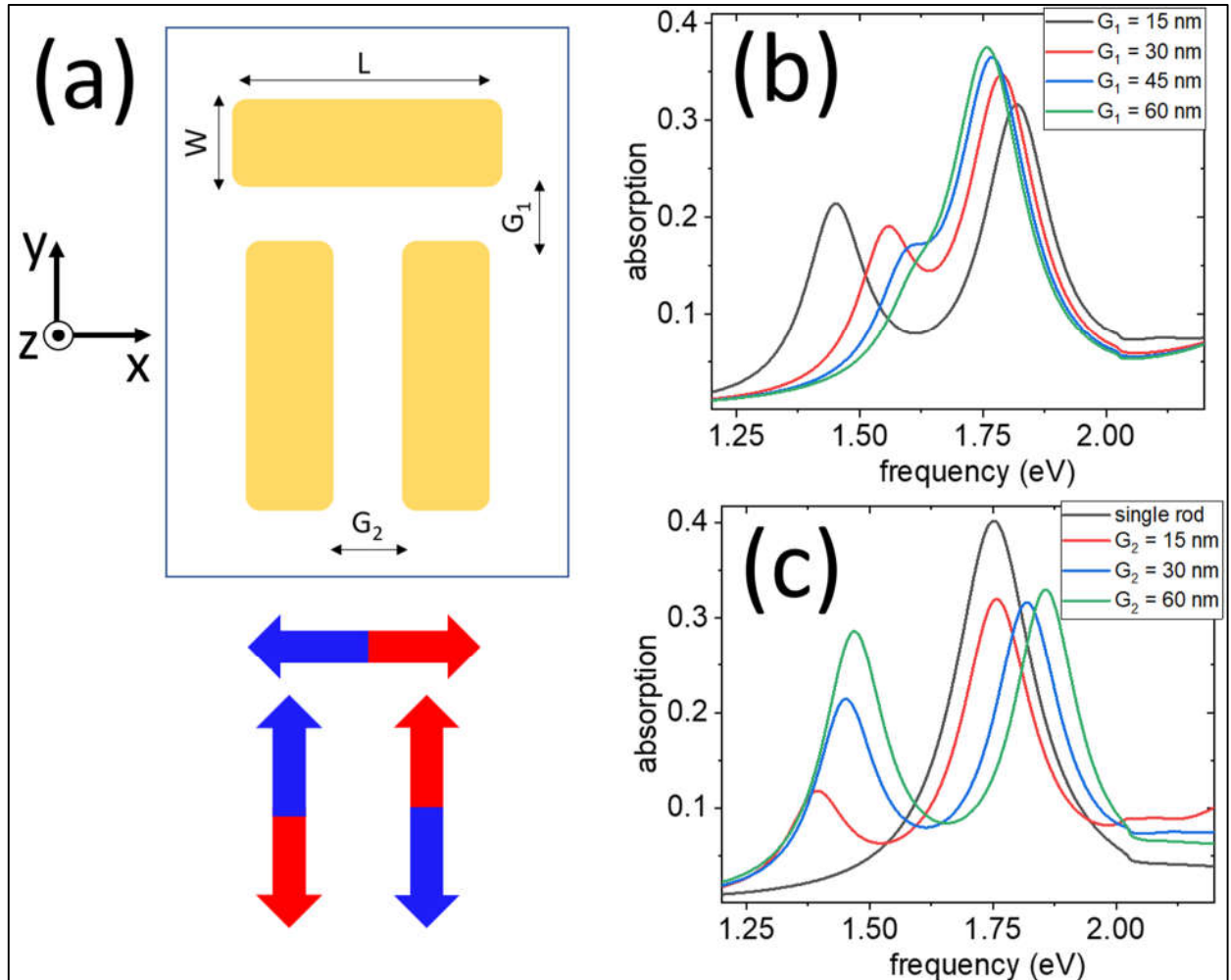

**Fig. 1**. Linear optical properties of a periodic array of nanodolmens. Panel (a) shows a schematic setup of a nanodolmen comprised of three nanorods. Blue-red arrows indicate interacting plasmon modes: bright mode (horizontal arrow) and dark mode (two vertical arrows). Panel (b) shows the linear absorption spectra when varying gap $G_1$: (black line) 15 nm, (red line) 30 nm, (blue line) 45 nm, (green line) 60 nm. Parameter $G_2$ is 30 nm. Panel (c) shows absorption spectra for different $G_2$: (black line) only horizontal nanorod is considered, (red line) 15 nm, (blue line) 30 nm, (green line) 60 nm. $G_1$ is 15 nm. Other parameters for both panels are: periodicity is 405 nm, $W = 40$ nm, $L = 110$ nm, and the thickness of each nanorod is 60 nm.

Figs. 1b and 1c summarize the main properties of the linear response of the dolmen arrays. Firstly, we note that the single nanorod has a mode centered at 1.75 eV (black line, Fig. 1c). Once the two parallel wires are placed nearby, the absorption exhibits two peaks, which correspond to the normal modes of the nanodolmen due to hybridization between the bright and the dark modes. Fig. 1b shows the linear absorption spectra for different values of the parameter $G_1$ that controls the near-field coupling strength between the bright and dark modes. One can see how sensitively the coupling depends on such a parameter. When it changes from 15 nm (black line) to 30 nm (red line), the splitting between the peaks (determined by the coupling strength) drops by over 60%. It continues to drop for higher values of $G_1$ and eventually becomes negligible at around 60 nm

(green line). Fig. 1c explores the linear absorption dependence on parameter $G_2$, which controls the frequency of the dark mode and its linewidth. The larger the gap gets the higher the frequency of the mode is, which in turn affects the coupling as well. Contrary to increasing $G_1$, higher values of the gap between two vertical wires, $G_2$, increases the splitting between the observed modes and also enhances the signal at the low frequency normal mode. Corresponding results for an individual dolmen are summarized in the Supplementary Information, including the linear and nonlinear response (Fig. S1).

We now turn to the second order nonlinear response of the nanodolmen array. The system is driven by a 500-fs pump with an amplitude of 0.05 V/nm. Power spectra on the input (reflection) and output (transmission) sides are calculated. Calculations of the second order susceptibility follow the numerical procedure outlined elsewhere.[31] The outgoing electromagnetic energy flux at the second harmonic of the pump frequency is calculated on both sides of the array and the hyperpolarizability per unit cell, $\beta$, is extracted. The second order susceptibility is evaluated according to

$$\chi^{(2)} = \frac{\beta}{2V}, \qquad (5)$$

where $V$ is the volume occupied by the nanodolmen. We note that since the outgoing flux is integrated over all orientations at detection planes placed in the far-field zone obtained values of $\chi^{(2)}$ correspond to the orientation average.

First, we examine the angular distributions of the second harmonic signal separating horizontal and vertical in-plane components of the signal. When in-plane polarization of the pump changes we can observe which component of the susceptibility tensor contributes the most to the signal.[34] With a detector placed sufficiently far away the second order polarization reads

$$P_i^{(2)} \sim \sum_{j,k=x,y} \chi_{ijk}^{(2)} E_j E_k, \qquad (6)$$

where $E_{j,k}$ are the in-plane components of the driving field. We note that one needs to carefully check if the detection is indeed performed in the far-field zone by comparing the functional dependence of the second order response on the polarization angle at different nanodolmen-to-detection plane distances. For the parameters considered we found that for distances over 400 nm numerical convergence is achieved.

Figs. 2a and 2b present the angular dependence of the second harmonic signal on the polarization angle $\theta$ (where $\theta = 0^0$ corresponds to the pump polarized along the upper nanorod aligned in x-direction). Simulations are performed for the nanodolmen with $G_1 = 15$ nm and $G_2 = 30$ nm (the rest of the parameters are the same as in Fig. 1). Fig. 2a shows the horizontally polarized second harmonic signal and Fig. 2b shows the vertically polarized signal. These are proportional to $P_{x,y}^{(2)}$, correspondingly. When the nanodolmen is driven at its lower frequency mode 1.45 eV (see Fig. 1c, blue line), the angular distribution plotted in blue clearly exhibits the following two dependencies

$$\begin{aligned} P_x^{(2)} &\sim \chi_{xxy}^{(2)} \sin(2\theta), \\ P_y^{(2)} &\sim \chi_{yxx}^{(2)} \cos^2(\theta), \end{aligned} \qquad (7)$$

indicating that the off-diagonal elements $\chi^{(2)}_{xxy}$ and $\chi^{(2)}_{yxx}$ are the primary sources of the second harmonic generation, which in turn means that the dark mode that mixes $x$ and $y$ polarizations contributes significantly to the optical response at that driving frequency. Contrary to such observations, when the system is driven at its higher frequency mode 1.82 eV (see Fig. 1c, blue line), the angular dependence of the second harmonic (shown in red) follows expected dipolar patterns

$$P^{(2)}_x \sim \chi^{(2)}_{xxx} \cos^2(\theta), \\ P^{(2)}_y \sim \chi^{(2)}_{yyy} \sin^2(\theta), \quad (8)$$

with diagonal elements of $\chi^{(2)}_{ijk}$ dominating the second harmonic signal.

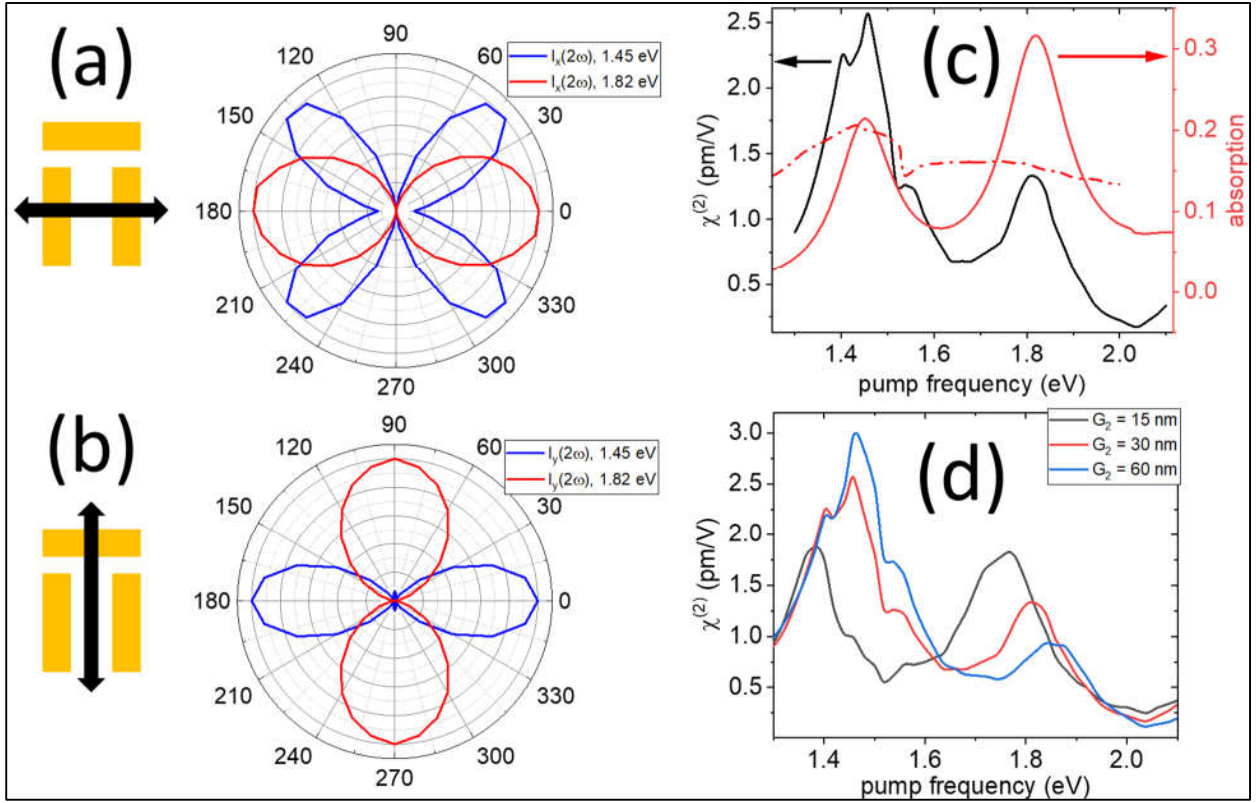

**Fig. 2**. Second harmonic generation by nanodolmens. Panels (a) and (b) show angular diagrams of the second harmonic signal polarized horizontally (panel (a)) and vertically (panel (b)) as function of the in-plane polarization angle of the pump. Simulations are performed for the nanodolmen with $G_1$ = 15 nm and $G_2$ = 30 nm with the other parameters as those in Fig. 1. Blue lines indicate results obtained at the driving frequency of 1.45 eV, corresponding to the antisymmetric normal mode, and red lines show data for the driving frequency 1.82 eV, i.e., the symmetric normal mode. Panel (c) compares the second order susceptibility with a linear absorption on a double vertical scale for the geometrical parameters as in panels (a) and (b). The solid red line shows absorption as a function of frequency and the dash-dotted line shows the same absorption plotted at half the frequency. The solid black line shows $\chi^{(2)}$ as a function of the pump frequency. Panel (d) shows the second order susceptibility as a function of the pump frequency for $G_2$ = 15 nm (black line), $G_2$ = 30 nm (red line), $G_2$ = 60 nm (blue line).

Fig. 2c explores the orientation average $\chi^{(2)}$ as a function of the pump frequency and compares it with the linear absorption plotted as a function of the driving frequency (solid red line), and half the frequency (dash-dotted red line) to explore two-photon processes due to the high frequency modes to $\chi^{(2)}$. First, we note that unlike the linear response exhibited by the nanodolmen, where the linear scattering signal is noticeably higher at 1.82 eV (blue line in Fig. 1c), the second order signal is dominated by the low frequency mode. Secondly, resonances at 1.4 eV and 1.53 eV exhibited by $\chi^{(2)}$ are due to the second harmonic contributions from the high energy quadrupole mode and Wood's anomaly corresponding to the periodicity of the nanodolmen array. The period used in the simulations is 405 nm, namely 2×1.53 eV.

We alter the coupling strength between the bright and dark modes by varying parameter $G_2$ and extract the orientation average $\chi^{(2)}$ as a function of the pump frequency. The calculated second order response shows significant enhancement at the lower frequency mode and the suppression of the signal at the high frequency mode. When the coupling between the bright and dark modes increases along with the frequency position of the dark mode, $\chi^{(2)}$ visibly decreases at near 1.8 eV and is significantly enhanced near 1.45 eV as seen in Fig. 2d. In part, the observed enhancement can be explained by the small damping of the low frequency mode as follows from the model of coupled oscillator model discussed in the next section. Additionally, it has been shown[25] that the local field enhancement for such a mode can reach several orders of magnitude, which in turn leads to more efficient second harmonic generation. The observed enhancement is quite general and should be exhibited by other plasmonic systems. In the Supplementary Information we provide a second example of a Fano-type plasmonic system, namely nano-pentamer[35] and examine its second order response. In brief, our simulations support main prediction – the second harmonic generation process driven at the frequency corresponding to a narrow resonant mode with a Fano-like profile leads to strong enhancement of the nonlinear response. We also analyze the behavior of the second harmonic signal generated by an individual nanodolmen, where small deviations in geometry are introduced. It is demonstrated that the angular properties of the second harmonic markedly depend on geometrical imperfections. However, the nonlinear signal enhancement due to local field enhancement is less sensitive and is still observed.

**Analytical model of two coupled nonlinear oscillators**

To elucidate the second-order nonlinear response from a Fano type nonlinear plasmonic system we consider a model of two coupled oscillators, both exhibiting second order nonlinearities.[36] This model is an extension of the widely used coupled oscillators model[37] that found a wide variety of applications in physics, materials science, and chemistry.[23] The toy model is schematically depicted in Fig. 3a. Here oscillator B is directly driven by the external force $F_0$. Oscillator A is coupled to B and can be excited only via B. We further assume that each oscillator has a second order nonlinearity. The equations of motion read

$$\begin{aligned}\ddot{x}_A + \gamma_A \dot{x}_A + \omega_A^2 x_A + \alpha_A x_A^2 + g^2 x_B &= 0, \\ \ddot{x}_B + \gamma_B \dot{x}_B + \omega_B^2 x_B + \alpha_B x_B^2 + g^2 x_A &= F_0\, e^{i\omega t},\end{aligned} \qquad (9)$$

Here $g$ is the coupling strength between A and B and $\alpha_{A,B}$ are the parameters characterizing the strength of nonlinearity. The analogy between the model and the nanodolmen is clear: the bright mode corresponding to the longitudinal plasmon mode of the top nanowire (see Fig. 1a) is the driven oscillator B, the dark mode having a higher frequency and lower damping rate is oscillator A.

The solution of Eqs. (9) can written as a perturbation series

$$x_{A,B} = \sum_{n} x_{A,B}^{(n)}, \quad (10)$$

where $x^{(n)} \sim F_0^n$. Inserting (10) into (9) and equating same perturbation orders we obtain a set of coupled differential equations of various orders

$$\begin{cases} \ddot{x}_A^{(1)} + \gamma_A \dot{x}_A^{(1)} + \omega_A^2 x_A^{(1)} + g^2 x_B^{(1)} = 0, \\ \ddot{x}_B^{(1)} + \gamma_B \dot{x}_B^{(1)} + \omega_B^2 x_B^{(1)} + g^2 x_A^{(1)} = F_0 e^{i\omega t}, \\ \ddot{x}_A^{(2)} + \gamma_A \dot{x}_A^{(2)} + \omega_A^2 x_A^{(2)} + \alpha_A \left(x_A^{(1)}\right)^2 + g^2 x_B^{(2)} = 0, \\ \ddot{x}_B^{(2)} + \gamma_B \dot{x}_B^{(2)} + \omega_B^2 x_B^{(2)} + \alpha_B \left(x_B^{(1)}\right)^2 + g^2 x_A^{(2)} = 0, \\ \ldots \end{cases} \quad (11)$$

Solutions of (11) corresponding to the first-order response (i.e. at the fundamental frequency of the driving force) are well-known and can be easily obtained by substituting $x^{(1)} \to x^{(1)} e^{i\omega t}$ and equating terms with $e^{i\omega t}$. This results in

$$\begin{aligned} x_A^{(1)} &= \frac{g^2}{g^4 - D_A(\omega) D_B(\omega)} F_0, \\ x_B^{(1)} &= -\frac{D_A(\omega)}{g^4 - D_A(\omega) D_B(\omega)} F_0, \end{aligned} \quad (12)$$

where the following notations are introduced $D_{A,B}(\omega) = \omega_{A,B}^2 - \omega^2 + i\gamma_{A,B}\omega$. The Fano-like lineshape can be readily obtained from (12) by assuming that the driven oscillator B has a higher damping rate compared to oscillator A, i.e. $\gamma_B \gg \gamma_A$. The interference that gives rise to the Fano-like resonance is due to the interference between the direct excitation of B by the driving force and the excitation of B via coupling through A. Fig. 3b shows $x_B^{(1)}$ as a function of the driving frequency at different coupling strengths, $g$. The Fano-like line shape becomes clearly visible at higher values of $g$. When $g$ becomes larger than $\gamma_A$ two normal modes corresponding to antisymmetric and symmetric combinations of $x_A^{(1)}$ and $x_B^{(1)}$, respectively, are formed at the frequencies

$$\Omega_{\mp} = \left( \frac{\omega_A^2 + \omega_B^2 + \gamma_A \gamma_B}{2} \mp \frac{1}{2} \sqrt{4g^4 + \left(\omega_A^2 - \omega_B^2\right)^2 + 2\gamma_A \gamma_B \left(\omega_A^2 + \omega_B^2\right) + \gamma_A^2 \gamma_B^2} \right)^{1/2}. \quad (13)$$

Here we keep the decay rates $\gamma_{A,B}$ when deriving (13) from equating the real part of the denominator of (12) to 0. Eq. (13) reduces to well-known expressions if $\gamma_{A,B}$ are neglected.[38]

Evidently, the second order displacement oscillates at the second harmonic of the driving force. Substituting $x^{(2)} \to x^{(2)} e^{2i\omega t}$ and inserting (12) into (11) leads to

$$x_A^{(2)}(2\omega) = \frac{\alpha_A g^4 D_B(2\omega) - \alpha_B g^2 D_A^2(\omega)}{\left(g^4 - D_A(\omega) D_B(\omega)\right)^2 \left(g^4 - D_A(2\omega) D_B(2\omega)\right)} F_0^2,$$

$$x_B^{(2)}(2\omega) = \frac{\alpha_B D_A^2(\omega) D_A(2\omega) - g^6 \alpha_A}{\left(g^4 - D_A(\omega) D_B(\omega)\right)^2 \left(g^4 - D_A(2\omega) D_B(2\omega)\right)} F_0^2.$$

(14)

We note that (14) is the response at the second harmonic of the driving force and has two distinct contributions from one-, $D(\omega)$, and two-photon, $D(2\omega)$, emission although the latter is usually appreciably weaker.[20]

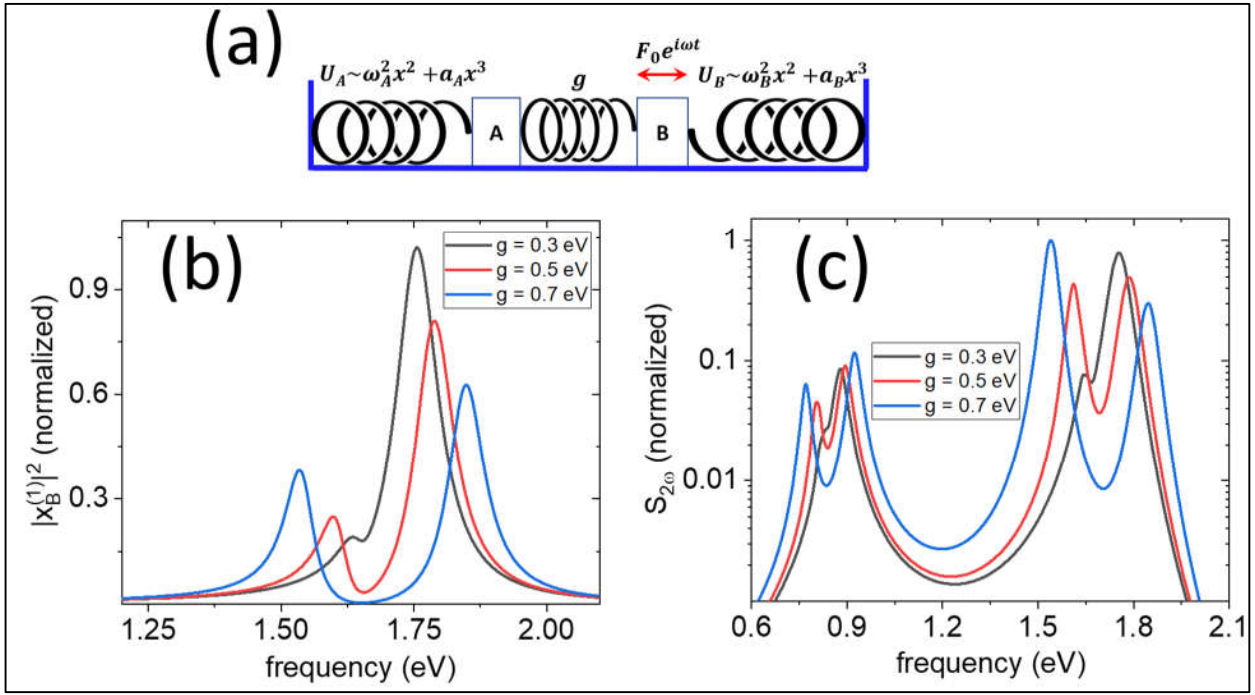

**Fig. 3**. Two coupled nonlinear oscillators. Panel (a) shows a schematic setup of the model, where oscillator $B$ is driven by the periodic force and oscillator $A$ coupled to $B$ with coupling strength $g$. Both oscillators have the third order nonlinearity in potential energy (i.e., the second order in the corresponding restoring force). Panel (b) shows the first order signal by oscillator $B$ as a function of the driving frequency for different values of the coupling strength $g$: (black line) $g = 0.3$ eV, (red line) $g = 0.5$ eV, (blue line) $g = 0.7$ eV. Panel (c) shows the second order signal defined in (15) as a function of the driving frequency for the same values of the coupling $g$. Other parameters are: $\omega_A = 1.65$ eV, $\gamma_A = 0.05$ eV, $\omega_B = 1.75$ eV, $\gamma_B = 0.1$ eV, $\alpha_A = \alpha_B = 10^{-3}$ eV$^2$/m.

Our primary interest is to investigate the second harmonic response pertaining to the coupled plasmon modes that form Fano-like resonances. The emission signal from the coupled oscillators $A$ and $B$ can be written as

$$S_{2\omega} \sim \left|x_A^{(2)}(2\omega)\right|^2 + \left|x_B^{(2)}(2\omega)\right|^2.$$

(15)

Fig. 3c explores $S_{2\omega}$ as a function of the driving frequency at different coupling strengths, $g$. It follows from (14) and (15) that when $g=0$ the second harmonic response from oscillator $A$ is strictly 0 by the virtue of the model since $A$ is only excited via coupling to $B$. $S_{2\omega}$ in such a situation expectedly exhibits two peaks at $\omega_B$ and $\omega_B/2$ (the latter is due to two-photon emission). The behavior of the emission drastically changes when the coupling is present as shown in Fig. 3c. Each peak for the case of $g=0$ splits into two resonances with the frequencies defined by equation (13). The main observation is that the antisymmetric mode, $\Omega_-$, clearly results in a significant enhancement of the second harmonic response, while the contribution to $S_{2\omega}$ by the symmetric mode at $\Omega_+$ is obviously becomes weaker with increasing coupling. Since oscillator $A$ has considerably lower losses compared to $B$, it can store higher amounts of energy, which in turn leads to stronger emission. It was shown analytically using the model of coupled harmonic oscillators that the antisymmetric mode, $\Omega_-$, can be thought of as a subradiant mode with a linewidth due to nonradiative losses only.[25] On the other hand, the symmetric mode has a larger width (superradiant mode). This in part explains the observed enhancement of the second harmonic in nanodolmens (Fig. 2d), where the calculated second harmonic signal is visibly enhanced at the low frequency mode (corresponding to the antisymmetric mode at $\Omega_-$ in the analytical model) and is suppressed at the high frequency mode ($\Omega_+$).

In summary, the proposed analytical model of coupled nonlinear oscillators predicts noticeable enhancement of the second order response at the frequency of the antisymmetric mode and the suppression of the signal by the symmetric mode. The primary source of the enhancement and suppression is the sub/superradiant nature of the normal modes of coupled oscillators.

**Conclusion**

Nonlinear optics of plasmonic systems exhibiting Fano resonances was investigated using both numerical solutions of the three-dimensional Maxwell-hydrodynamics equations and a simple model of coupled nonlinear oscillators. We investigated the second order nonlinear response by periodic arrays of nanodolmens that sustain Fano-like resonances in the absorption spectra. By employing the semi-classical hydrodynamics model for conduction electrons and coupling it directly to Maxwell's equations we numerically demonstrated the enhancement of the second harmonic generation by the mode with a small damping. Angular dependencies of the second harmonic response on the pump polarization driven at the frequency of this mode were shown to be dominated by the off-diagonal elements of the second order susceptibility. When the coupling strength between the bright and dark modes of the nanodolmen was increased the second harmonic generation signal was observed to be enhanced when driven at $\Omega_-$. On the other hand, when we drove the system at $\Omega_+$ the signal was suppressed. The proposed analytical model confirmed the significant enhancement of the second order response at a frequency, $\Omega_-$, corresponding to the antisymmetric normal mode. We also simulated second order susceptibilities as a function of the pump frequency for nanopentamers. The results of these simulations are collected and described in detail in Supplementary Information. Additionally, it was shown that the angular dependence of the second harmonic generated by nanodolmens varied sensitively with small spatial deviations of the geometry from its ideal shape, as summarized in the Supplementary Information.

**Acknowledgements**

This work is supported by the Air Force Office of Scientific Research under Grant No. FA9550-22-1-0175. The computational part of the work has been made possible via the Department of Defense High Performance Computing Modernization Program.

**Data availability**

The data that support the findings of this study are available from the corresponding author upon reasonable request.

# Supplementary Information: Fano plasmonics goes nonlinear

Maxim Sukharev, Elena Drobnyh, Ruth Pachter

**Numerical procedure to calculate hyperpolarizabilities of plasmonic systems**

The proposed approach consists of calculating scattering cross-section defined as[1]

$$C_{sca} = \frac{1}{I_{inc}} \oiint_A \vec{S}_{sca} \cdot d\vec{A}, \tag{1}$$

where $I_{inc}$ is the incident irradiance and the numerator corresponds to the outgoing electromagnetic (EM) energy flux integrated over a Gaussian surface, $A$, enclosing a nanostructure under consideration, $\vec{S}_{sca}$ is the Poynting vector comprised of scattered EM field.[2] Assuming that the Gaussian surface is in the far-field zone one can approximate the nanostructure as a point dipole. The linear response at the fundamental frequency can then be written as

$$p(\omega) = \varepsilon_0 \varepsilon_m \alpha(\omega) E_0(\omega), \tag{2}$$

here $\varepsilon_m$ is the relative permittivity of the media, in which the nanostructure is embedded, $\alpha$ is the polarizability, and $E_0$ is the peak amplitude of the pump. Combining (1), (2), and employing analytical solution for a Hertzian dipole[3] the scattering cross-section reads[1]

$$C_{sca}(\omega) = \frac{\omega^4}{6\pi c^4} |\alpha(\omega)|^2. \tag{3}$$

We thus can extract absolute value of the polarizability directly from numerical integration of Maxwell's equations coupled to equations governing dynamics of conduction electrons in metal. We note that the imaginary part of the polarizability can also be calculated from the extinction cross-section, which has the form

$$C_{ext}(\omega) = \frac{\omega}{c} \text{Im}(\alpha(\omega)). \tag{4}$$

We now proceed along the same lines and consider a dipole moment written as a Fourier sum over harmonics of the incident field[4]

$$p = p_1(\omega) e^{i\omega t} + p_2(2\omega) e^{2i\omega t} + p_3(3\omega) e^{3i\omega t} + \ldots \tag{5}$$

The linear term corresponds to (2). The second and third order terms read[5]

$$p_2(2\omega) = \varepsilon_0 \varepsilon_m \frac{1}{2} \beta(2\omega) : E_0(\omega) E_0(\omega),$$

$$p_3(3\omega) = \varepsilon_0 \varepsilon_m \frac{1}{6} \gamma(3\omega) : E_0(\omega) E_0(\omega) E_0(\omega), \tag{6}$$

where $\beta$ and $\gamma$ are hyperpolarizabilities (in molecular spectroscopy they are frequently referred to as first and second hyperpolarizabilities, respectively). Following the analogy with (3) the scattering cross-section calculated at the second harmonic of the pump is

$$C_{sca}(2\omega) = \frac{2\omega^4}{3\pi c^4} \cdot |\beta(2\omega)|^2 \cdot E_0^2. \tag{7}$$

Direct calculations of the scattering cross-section at the second harmonic and subsequent comparison with (7) allows us to directly extract hyperpolarizabilities. This approach has been recently applied to calculate hyperpolarizabilities of triangular shaped nanoprisms[6]. Numerical data agreed very well with measurements.

This procedure can be extended to calculate hypolarizabilities of periodic arrays (in this case calculated hypolarizability is per unit cell). The scattering cross-section is replaced with the signal detected on the input (reflection) and output (transmission) sides with the rest of simulations analogous to the (7).

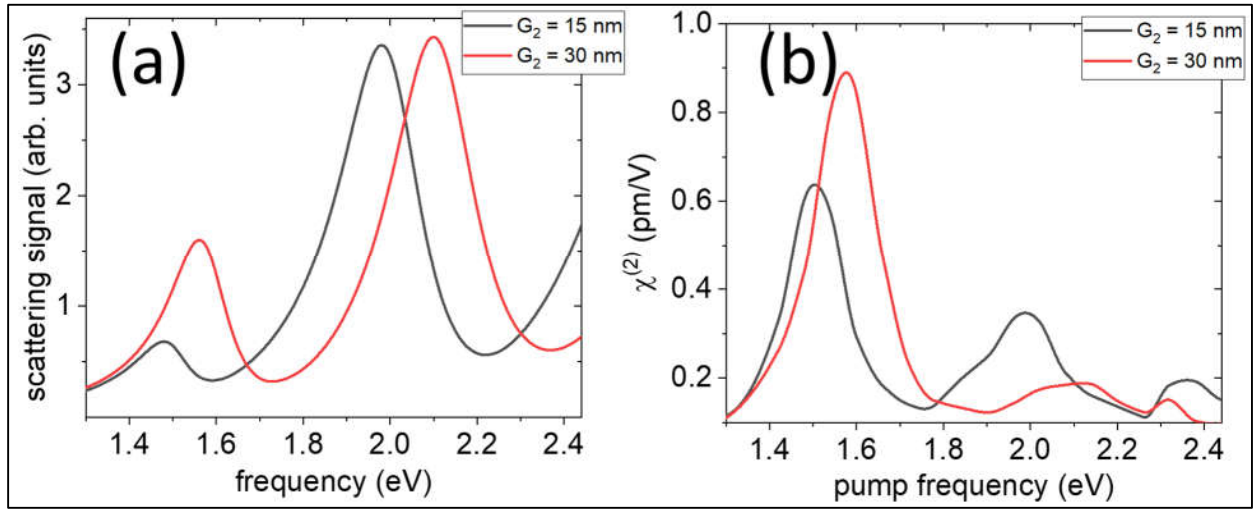

**Fig. S1**. Stand-alone nanodolmen simulations. Panel (a) shows linear scattering signal as a function of frequency for Au nanodolmen with $G_2$: (black line) 15 nm, (red line) 30 nm. The rest of the parameters for both panels are: $G_1$ is 15 nm, $W = 40$ nm, $L = 110$ nm, thickness of each nanorod is 60 nm. See Fig. 1a for details of the geometry.

**Second harmonic generation by stand-alone nanodolmens**

In the manuscript we considered periodic arrays of nanodolmens with a unit cell schematically depicted in Fig. 1a primarily due to easier experimental implementation. Single nanodolmens exhibit optical response similar to periodic arrays as shown in Fig. S1. When parameter $G_2$ is varying analogous to Fig. 1c, the antisymmetric mode of the nanodolmen can be altered. This in turn leads to second harmonic enhancement as shown in Fig. S1b. The low frequency pump enhances the signal while the high energy mode suppresses second order response as discussed in the manuscript.

Simulations of a single dolmen allows us to explore small deviations of the geometry from its ideal shape. Fig. S2 summarizes our results showing angular diagrams of the second harmonic response calculated when the nanodolmen is driven by the pump resonant with the antisymmetric mode. Each panel shows second harmonic response detected in x- (blue) and y-direction (red) (see Fig. 1a that defines our system of coordinates) as a function of the in-plane incident angle.

Firstly, we note that the quadrupole nature of the dark mode is less pronounced in case of a single nanodolmen compared to the periodic array – compare Fig. 2b (blue) with the upper left panel of Fig. S2 (blue). Although the angular dependence $\sin(2\theta)$ is seen in both cases, stand-alone nanodolmen also has clear contribution $\cos^2(\theta)$ making the diagram more elongated along

0-180⁰. This in turn indicates that $\chi^{(2)}_{xxx}$ plays significant role. In case of periodic arrays such a contribution is negligibly small making the second harmonic enhancement at the antisymmetric mode even more pronounced.

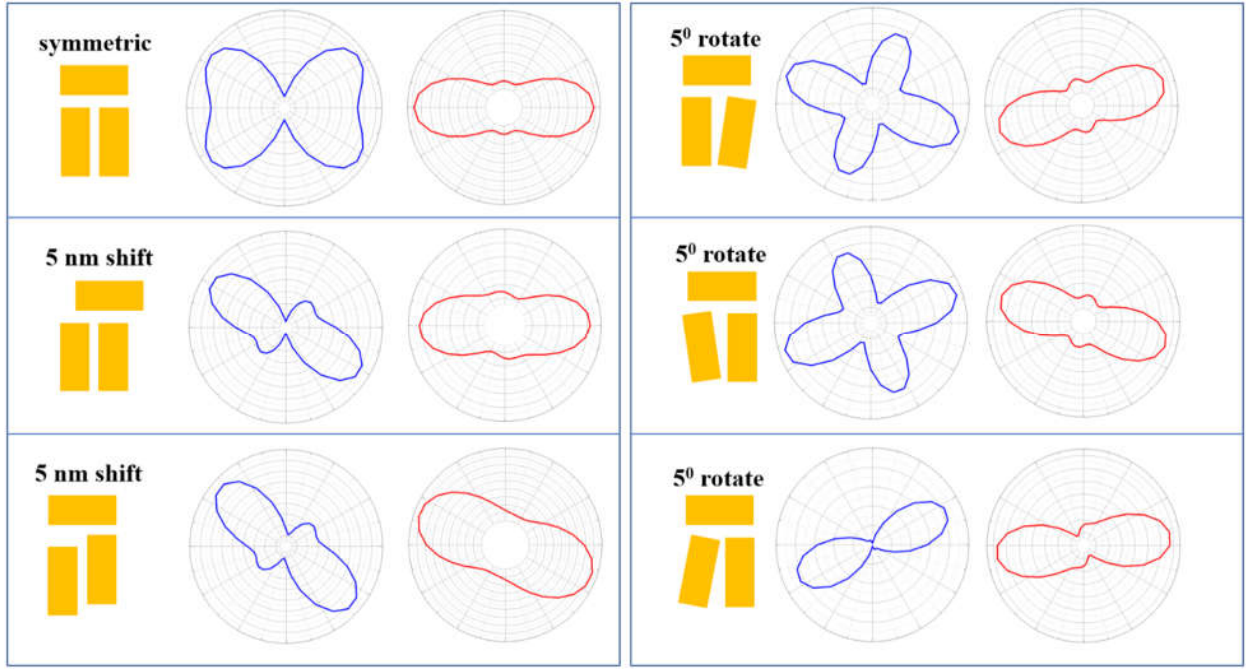

**Figure S2**. Second harmonic signal detected in *x*- (blue) and *y*-directions (red) as a function of in-plane incident angle. Each panel is simulated when driving the nanodolmen at 1.5 eV corresponding to the antisymmetric mode (see Fig. S1a, black line). $G_2$ = 15 nm and the rest of the parameters are the same as in Fig. S1.

When small spatial deviations are introduced, the angular diagrams change dramatically. We note however that horizontal and vertical shifts of nanorods (Fig. S2, left column, panels in the middle and at the bottom) the four-fold angular diagrams are significantly distorted. This is due to the fact that the coupling between the bright and dark modes is greatly skewed. Rotating nanorods counterclockwise (Fig. S2, right column, upper panel, and panel in the middle) does influence the response but maintains four-fold functional dependence. Unlike the latter rotations of the left nanorod clockwise by only $5^0$ nearly completely destroys the four-fold picture. Interestingly such spatial deviations barely affect linear scattering response, but the second order signal is highly sensitive to various changes in geometry.

**Second harmonic generation by nanopentamers**
To demonstrate the generality of the second harmonic enhancement by Fano-type plasmonic systems we also consider a nanopentamer schematically shown in Fig. S3a comprising six Au nanoprisms. It has been shown that systems with various arrangements of nanoprisms may exhibit Fano response.[7] We consider a single pentamer with a nanoprism of larger radius placed in the center and surrounded by five nanoprisms of a smaller radius. The Fano response is observed due to the near-field coupling between two modes: collective dipolar mode from five small nanoprisms and the dipolar mode of the central nanoprism. Since both modes can be excited directly by the pump the Fano profile appears (unlike in nanodolmens) due to the difference between decay rates of two modes. The symmetric mode in this case has a Fano lineshape and thus leads to the second harmonic enhancement. Our results are summarized in Fig. S3. Here we observe enhancement of the second harmonic signal at the frequency corresponding to the resonance seen in the linear spectrum (red line) with a distinct Fano profile. Angular diagrams of the second harmonic response also indicate the importance of the off-diagonal elements of $\chi^{(2)}$ when the system is driven at the Fano-type resonance. We note that similar results can be obtained by having a small nanoprism in the center and surrounding it with large particles. In this case the antisymmetric mode exhibits the Fano profile.

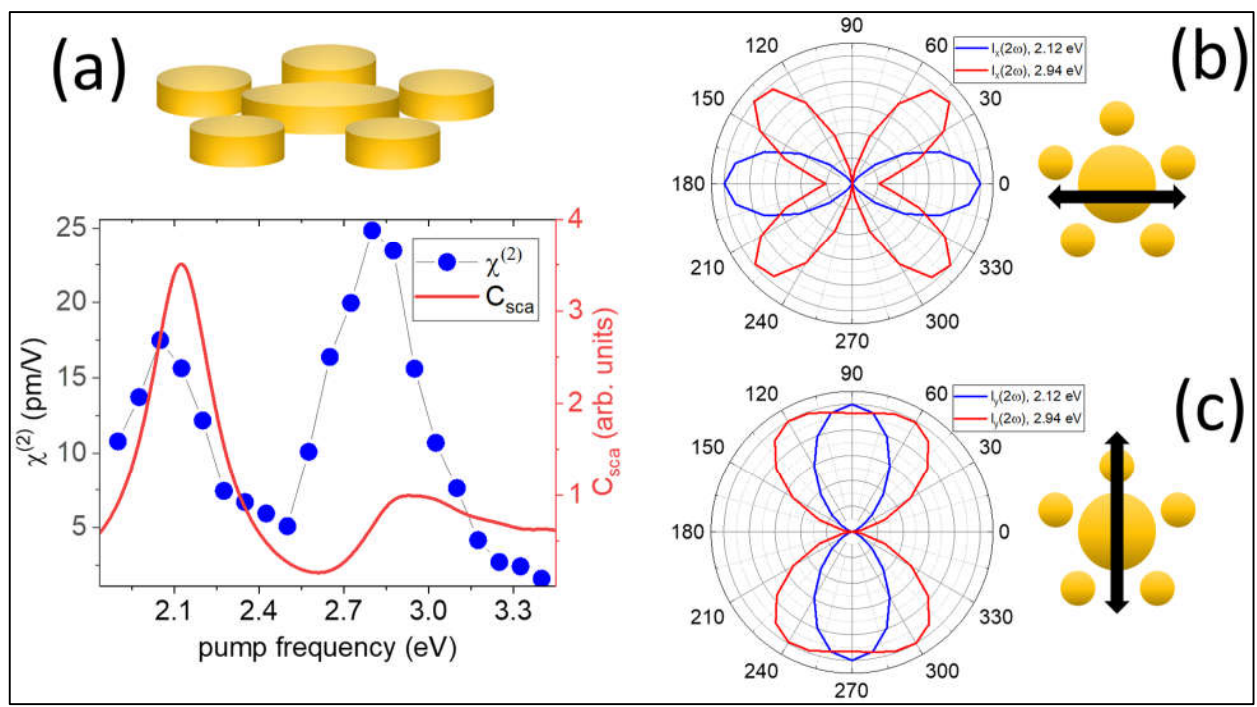

**Fig. S3**. Optics of Au nanopentamers. Nanopentamer is shown schematically in the upper part of panel (a). Each nanoprism is 20 nm high. Small nanoprisms with a radius of 20 nm are placed equidistantly along a circle with a radius of 70 nm. The radius of the nanoprism in the center is 40 nm. Panel (a) shows linear scattering signal (red line, left vertical axis) and the second-order susceptibility as functions of the pump frequency. Panels (b) and (c) show angular diagrams of the second harmonic response when driven at 2.12 eV (blue) and 2.94 eV (red). The latter corresponds to the mode with a Fano profile. Panel (b) shows horizontally polarized signal and panel (c) shows vertically polarized signal as indicated by two cartoons on the right of each angular diagram.